\documentclass[preprint,superscriptaddress,aps,pre]{revtex4}
\usepackage{graphicx,amsmath}

\usepackage{graphicx}

\date{\today}

\begin{document}

\title{
Onsager-Machlup action-based path sampling
and its combination with replica exchange 
for diffusive and multiple pathways
}

\author{Hiroshi Fujisaki}\email{fujisaki@nms.ac.jp} 
\affiliation{
Department of Physics,
Nippon Medical School,
2-297-2 Kosugi-cho, Nakahara, Kawasaki 
211-0063, Japan
}
\affiliation{
Molecular Scale Team, 
Integrated Simulation of Living Matter Group,
Computational Science Research Program,
RIKEN, 2-1 Hirosawa, Wako 351-0198, Japan
}
\author{Motoyuki Shiga}\email{shiga.motoyuki@jaea.go.jp} 
\affiliation{
Center for Computational Science and E-Systems,
Japan Atomic Energy Agency (JAEA),
Higashi-Ueno 6-9-3, Taito-ku, Tokyo 110-0015, Japan
}
\author{Akinori Kidera}\email{kidera@tsurumi.yokohama-cu.ac.jp} 
\affiliation{
Molecular Scale Team, 
Integrated Simulation of Living Matter Group,
Computational Science Research Program,
RIKEN, 2-1 Hirosawa, Wako 351-0198, Japan
}
\affiliation{
Department of Supramolecular Biology, Graduate School of Nanobioscience,
Yokohama City University, 1-7-29 Suehiro-cho, Tsurumi, 
Yokohama 230-0045, Japan
}

\begin{abstract}




For sampling multiple pathways in a rugged energy landscape,
we propose a novel action-based path sampling method using
the Onsager-Machlup action functional.
Inspired by the Fourier-path integral simulation of a quantum mechanical
system, 
a path in Cartesian space is transformed into that in Fourier space,
and an overdamped Langevin equation is derived for 
the Fourier components to achieve a canonical ensemble of the path 
at a finite temperature.
To avoid ``path trapping'' around an initially guessed path,
the path sampling method is further combined with 
a powerful sampling technique, the replica exchange method.
The principle and algorithm of our method is numerically demonstrated for
a model two-dimensional system with a bifurcated potential landscape.
The results are compared with those of conventional 
transition path sampling 
and the equilibrium theory, and the error due to path discretization 
is also discussed. 

\end{abstract}





\maketitle

\section{Introduction}

To understand the functions of biomolecules (chemical reactions, conformational change, 
ligand-binding, protein-protein association etc.), determining reaction paths
is an essential step \cite{FFMK09}. 
However, conventional molecular dynamics (MD) simulations often 
fail to find appropriate reaction paths even 
with huge computational resources \cite{Shaw09}. 
This is 
because such reactions cannot be simulated within 
computationally feasible times, and 
the sampling efficiency of the reaction can be 
low especially for large systems. 
Handling such a rare event is a common issue 
not only for biomolecular systems such as proteins 
but also for many physical or chemical systems.


So far, many methodologies have been devised for
path search or path sampling in biomolecular systems \cite{FFMK09}.
The most primitive but powerful method may be
umbrella sampling after defining certain 
reaction coordinates
such as the radius of gyration or the number of native contacts 
\cite{TV77,Roux01}.
Alternatively, metadynamics has been used to compute 
free energy surfaces with several reaction coordinates \cite{LP02}.
However, all these methods only work for molecular processes 
characterized by limited numbers of reaction coordinates
that are known {\it a priori}, and one often fails to 
describe complex and highly multi-dimensional processes in 
biomolecular systems.
As the computational effort grows exponentially with
the number of reaction coordinates (dimensions) increases,
it is hopeless to work on the full detail of a high-dimensional
free energy landscape.
Steered or targeted molecular dynamics is another 
powerful and well-established path search method \cite{SMD,Vaart06}.
In combination with the Jarzynski equality \cite{Jarzynski97,Crooks98,PS04}, 
a free energy difference can be calculated by steered MD.
However, 
since the reaction path obtained only follows a 
predefined direction, it may sometimes result in producing paths
with unphysically large deformations of molecules.

More direct approaches to deal with complex multi-dimensional
reactions are based on minimization of a physically defined 
{\it action functional} associated with a reaction path.
Elber and Karplus \cite{EK87} proposed an innovative approach 
to use a line integral 
along a path with some constraints, in search of 
a minimum energy path such as the intrinsic 
reaction path \cite{Fukui81,Muller80,Shiga09}.
This method was successfully applied to peptides \cite{CE90} and water 
systems \cite{Ohmine}.
Other related methods searching for a minimum energy path 
include the nudged elastic band method \cite{NEB,Case},
the conjugate peak refinement method \cite{CPR,Fischer},
and the zero-temperature string method \cite{ERE07}.
These methods can be used to locate transition state geometries
(saddle points of the potential energy surface,
at which barrier crossing can occur 
with a minimum amount of activation energy),
but 
%
they do not contain any dynamical information by themselves, since
the transition state geometry is merely a static feature of
the potential energy landscape.
Some kinetic information is retrieved with the use of
the discrete path sampling method \cite{Wales02} under
the assumptions of transition state theory \cite{TST},
but the effect of thermal fluctuations,
which should be particularly important for biomolecular systems,
is not completely captured by these approaches.
To include dynamical and thermal effects,
transition path sampling (TPS) has been proposed 
and developed by
Chandler and coworkers \cite{TPS,TPS2}.
The basic idea of TPS is that a system is described by an ensemble of paths
instead of a single path.
(For a similar and heuristic approach, see \cite{SGD08}.)
When an overdamped Langevin equation is assumed to describe 
the dynamics, the weight for a path can be written 
as 
\begin{equation}
P[x(t)]
\propto 
e^{-S[x(t)]/2D}
\end{equation}
where 
\begin{equation}
S[x(t)]
=
\frac{\Delta U}{\zeta}
+\frac{1}{2}\int_0^t ds 
\left[ \dot{x}^2+\left(\frac{1}{\zeta} \frac{dU}{dx}\right)^2
-\frac{2D}{\zeta} \frac{d^2U}{dx^2}
\right]
\label{eq:OM}
\end{equation}
is called the 
Onsager-Machlup (OM) action 
``functional'' \cite{OM53,Wiegelbook}
(for its derivation and pedagogic discussions, see, 
for example, \cite{Adib08})
because it contains all the information of the whole history of a path $x(t)$.
Here, $\Delta U=U(x_{\rm fin})-U(x_{\rm ini})$ is the potential 
energy difference between an initial state 
with $x_{\rm ini}=x(0)$ and a final state with $x_{\rm fin}=x(t)$,
$D$ is the diffusion constant, which is related to 
a friction constant $\zeta$ and the absolute temperature $T$ 
by Einstein's relation $D=k_B T/\zeta$ ($k_B$ is the Boltzmann constant).
TPS intends to sample many paths according 
to the statistical weight above. 
Combining conventional MD simulations and Monte Carlo (MC) moves,
shooting and shifting \cite{TPS},
TPS has been successfully used for {\it rare} and {\it fast} chemical reactions.

However, in biomolecular reactions, such as folding 
or conformational changes, the process is often not only 
{\it rare} but also {\it slow} and {\it diffusive}. 
Protein folding can occur 
at least more than microsecond 
timescales, 
which is 
still beyond the ability of current ``brute-force'' MD simulations 
using ``standard'' computers and algorithms.
This is also the case for conformational change of large protein 
systems with several domains, which is relevant to 
the understanding of the 
functions of proteins \cite{FFMK09}.
Therefore, alternative approaches 
to sample path ensembles 
should be established,  
and novel hardware strategy \cite{Shaw09} and novel theoretical 
algorithms including transition interface sampling \cite{JB08},
milestoning \cite{Elber07}, and 
Markov state models \cite{Pande09} 
have been developed in the last decade.

In this paper, we pursuit another approach
by discretizing the OM action
\begin{equation}
S=
\frac{\Delta U}{\zeta}
+\sum_{i=1}^{N-1}
\left[ \frac{\kappa}{2} (x_{i+1}-x_i)^2 
+W(x_i)
\right] 
\end{equation}
where $\kappa=1/\Delta t$, $x_1=x_{\rm ini}, x_N=x_{\rm fin}$ and 
\begin{equation}
W(x)
\equiv 
\frac{1}{2 \kappa}
\left[ \left(\frac{1}{\zeta} \frac{dU}{dx}\right)^2
-\frac{2D}{\zeta} \frac{d^2U}{dx^2}
\right].
\end{equation}
The statistical weight is $P(x_1,x_2, \cdots, x_N) \propto e^{-S/2D}$,
and the path search problem is therefore mapped onto 
the statistical mechanics of a polymer under the effective potential 
function $W(x)$ \cite{Wiegelbook}. 
(This isomorphology is similar to that between a quantum particle 
and a ring polymer \cite{CW81}.)
This strategy was proposed in the original TPS 
paper \cite{DBCC98} but has not been completely worked out.
The previous studies using this type of action-based method have mostly focused 
on a {\it single} most probable path during conformational changes 
of biomolecules \cite{Elber,ZW99,EGD01,Orland09}. 
Here, we try to complete the previous studies  
and provide a path ensemble at finite temperatures using the 
action-based method,
which should be relevant for slow and diffusive 
biomolecular reactions.

To generate a path ensemble, we need to assume 
an initial path, from which the path sampling simulation starts.
If there is a barrier in path space, however, 
the generated path cannot move to the most probable path
because ``path trapping'' occurs in the basin around the initial path 
(as in the case of ``configuration
trapping'' for the folding problem of proteins).
Even worse, for complex systems with rugged energy
surfaces, the path to be calculated may not be unique,
and multiple paths can coexist.
This is the situation we want to address in this paper.
In the previous study using the nudged-elastic band,
simulated annealing was employed to avoid this 
``path trapping'' problem \cite{Case}.
This is a nice strategy to sample multiple 
{\it minimum energy paths}, however, a path ensemble 
{\it at a finite temperature} cannot be obtained. 
To generate a canonical ensemble for a path, 
we propose to introduce one of the generalized ensemble 
methods, the replica exchange method (REM) \cite{Hansmann97,SO99}, 
{\it in path space}.
The combination of REM with transition path sampling 
has been already proposed \cite{VS01,Frenkelbook,Bolhuis08}
for rare and fast chemical reactions, 
however, our aim is to sample rare, slow, and diffusive processes,
and the application of the OM action formalism should be more appropriate.

This paper is organized as follows.
In Sec.~\ref{sec:methods}, 
%
we derive an overdamped Langevin equation 
for a path such that
the path ensemble is generated according to the weight 
determined by the OM action.
In parallel to the technique employed in the path integral 
simulations of quantum systems, 
we use Fourier components of a path 
as ``dynamic'' variables \cite{CDF94}.
We then combine this method with REM
to improve the sampling efficiency in path space.
In Sec.~\ref{sec:result}, 
we test the numerical performance of our new method.
To this end, a two-dimensional model potential due to Bolhuis
\cite{Bolhuis08}, involving bifurcated reaction pathways, 
is numerically examined in detail.
The results are finally compared with those of TPS
and the equilibrium theory.
In Sec.~\ref{sec:summary}, 
we summarize the paper and discuss the 
connection between our method and other related ones
and future development for biomolecules.

\section{Theory}
\label{sec:methods}

\subsection{Onsager-Machlup action for multi-dimensional systems}

To describe slow and diffusive processes,
we start from an overdamped Langevin equation 
for an $M$-dimensional system using mass-weighted 
Cartesian coordinates $(x_1, \cdots, x_M)$:
\begin{equation}
\dot{x}_{\alpha}=-\frac{1}{\gamma}_{\alpha}\frac{\partial U}{\partial x_{\alpha}}
+\sqrt{2D_{\alpha}} \eta_{\alpha}(t)
\label{eq:Langevin}
\end{equation}
where $\gamma_{\alpha}$ is an intrinsic friction coefficient,
$\eta_{\alpha}(t)$ is a Gaussian-white noise satisfying 
$\langle \eta_{\alpha}(t)\eta_{\alpha'}(t') \rangle =\delta_{\alpha \alpha'}\delta (t-t')$, and 
\begin{equation}
D_{\alpha}=\frac{k_B T}{\gamma_{\alpha}} 
\label{eq:Einstein}
\end{equation}
is imposed by the fluctuation-dissipation theorem.
The corresponding Fokker-Planck equation \cite{FP} is 
\begin{equation}
\frac{\partial P(\{ x_{\alpha'} \},t)}{\partial t}
=\sum_{\alpha} \frac{\partial}{\partial x_{\alpha}}
\left( \frac{1}{\gamma_{\alpha}} \frac{\partial U}{\partial x_{\alpha}} 
P(\{ x_{\alpha'} \},t) \right)
+\sum_{\alpha} D_{\alpha} \frac{\partial^2}{\partial x_{\alpha}^2}
P( \{ x_{\alpha'} \},t).
\end{equation}
The path-integral representation of the propagator 
(Green's function) is written as \cite{Adib08,Orland09} 
\begin{equation}
P(x_{\rm fin}|x_{\rm ini};t) \propto
\int_{x(0)=x_{\rm ini}}^{x(t)=x_{\rm fin}} {\cal D}x(s) e^{- {\cal H}[x(s)]/k_B T},
\end{equation}
where ${\cal H}$ is the OM action 
defined by the multidimensional extension of Eq.~(\ref{eq:OM}),
multiplied by $\gamma_{\alpha}/2$ and summed over $\alpha$. 
Its discretized form is written as 
%
\begin{equation}
{\cal H}=
\frac{\Delta U}{2}
+\sum_{i=1}^{N-1}
\left[ 
\sum_{\alpha=1}^{M}
\frac{\omega_{\alpha}^2}{2} (x_{\alpha,i+1}-x_{\alpha,i})^2 
+V_{\rm eff}( \{ x_{\alpha,i} \} )
\right], 
\label{eq:OMHam}
\end{equation}
where $\Delta U = U(x_{\rm fin})-U(x_{\rm ini})$ and the effective potential is 
defined as 
\begin{equation}
V_{\rm eff}( \{ x_{\alpha} \})
\equiv 
\sum_{\alpha=1}^{M}
\left[ \frac{1}{8 \omega_{\alpha}^2}
U_{x_{\alpha}}^2
-\frac{k_B T}{4 \omega_{\alpha}^2}
 U_{x_{\alpha} x_{\alpha}}
\right]
\label{eq:veff}
\end{equation}
with 
\begin{equation}
\omega_{\alpha} = \sqrt{\frac{\gamma_{\alpha}}{2 \Delta t}}
\label{eq:omega}
\end{equation}
being an effective frequency determined by the friction and 
the time interval for the path discretization $\Delta t$.
(Hereafter $U_x, U_{xx}$ represent the first and second derivatives of 
$U$ with respect to $x$.)
Note that the path ensemble generated by the above OM ``Hamiltonian'' 
is isomorphic to the canonical ensemble of an $M \times (N-2)$ dimensional
(polymer) system. 


\subsection{Path sampling in Fourier space}

For simplicity, we first consider a one-dimensional system.
According to Cho, Doll, and Freeman \cite{CDF94}, 
the Fourier transform of a path 
connecting 
$x_1$ and $x_N$ 
is defined as 
\begin{equation}
x_i=x_i^{(0)}+ \sqrt{\frac{2}{N-1}} 
\sum_{k=2}^{N-1} q_k \sin \left( \frac{\pi (i-1)(k-1)}{N-1} \right)
=x_i^{(0)}+ \sum_{k=2}^{N-1} q_k u_{ik},
\end{equation}
where $q_k$ is the Fourier component of the path $x_i$, and 
$x_i^{(0)}$ is a reference path, which may be an initially guessed 
path.
To generate a path ensemble with 
the weight determined by the OM Hamiltonian ${\cal H}$, 
Eq.~(\ref{eq:OMHam}),
the following overdamped Langevin 
dynamics can be employed
\begin{equation}
\dot{q}_k 
= 
-L_k \frac{\partial}{\partial q_k} (\beta_T {\cal H})
+\sqrt{2 L_k} \eta_k(t),
\label{eq:odl}
\end{equation}
with 
\begin{equation}
\langle \eta_k(t) \eta_l(t') \rangle 
= 
\delta_{kl} \delta(t-t'),
\end{equation}
where $L_k$ is an Onsager coefficient representing friction,
and $1/k_B \beta_T$ is the temperature of a thermostat, 
not necessarily equal to the system temperature $T$. 
Using the corresponding Fokker-Planck equation,
the stationary point is shown to be 
the equilibrium state $P_{\rm eq}( \{ q_k \} ) \propto
\exp(-\beta_T {\cal H})$ in (discretized) path space.
Note that this Langevin dynamics is different from the original 
Langevin dynamics, Eq.~(\ref{eq:Langevin}), which was motivated from 
a physical consideration.

Substituting the 
particular form of the OM Hamiltonian,
Eq.~(\ref{eq:OMHam}),
into the overdamped Langevin equation,
we have 
\begin{eqnarray}
\dot{q}_k 
&=& 
-\beta_T L_k \sum_{i=2}^{N-1} u_{ik} \frac{\partial {\cal H}}{\partial x_i}
+\sqrt{2 L_k} \eta_k(t),
\\
\frac{\partial {\cal H}}{\partial x_i} &=&
- \omega^2(x_{i+1}+x_{i-1}-2x_i)+
\frac{\partial V_{\rm eff}(x_i)}{\partial x_i}, 
\\
\frac{\partial V_{\rm eff}(x)}{\partial x}
&=& 
\frac{1}{4 \omega^2}
U_x U_{xx}
-\frac{k_B T}{4 \omega^2} U_{xxx},
\end{eqnarray}
where $\omega$ is the one-dimensional analog of Eq.~(\ref{eq:omega}).
Finally we obtain
\begin{eqnarray}
\dot{q}_k &=&
G_k -\Gamma_k q_k 
-\beta_T L_k \sum_{i=2}^{N-1} u_{ik} 
\frac{\partial V_{\rm eff}(x_i )}{\partial x_i} 
+\sqrt{2 L_k} \eta_k(t),
\label{eq:langevin}
\end{eqnarray}
where 
\begin{eqnarray}
G_k &=& \omega^2 \beta_T L_k 
\sum_{i=2}^{N-1} 
u_{ik}(x^{(0)}_{i+1}+x^{(0)}_{i-1}-2x^{(0)}_i), 
\\
\Gamma_k
&=& 2 \omega^2 \beta_T L_k
\left( 1- \cos \frac{\pi (k-1)}{N-1} \right).
\end{eqnarray}
By taking 
\begin{equation}
L_k =\frac{1}{\lambda \Delta t \beta_T \omega^2}
\left( 1- \cos \frac{\pi (k-1)}{N-1} \right)^{-1},
\label{eq:Lk}
\end{equation}
the rate becomes constant: $\Gamma_k=\Gamma_0 
 = 2/(\lambda \Delta t)$
where $\lambda$ is a dimensionless parameter to adjust a 
timescale for relaxation. 
This particular form of the Onsager coefficient was chosen 
because the time step to solve the above 
series ($k=2,3, \cdots, N-1$) of Langevin 
equations may be determined by a single timescale $1/\Gamma_0$.

The Fourier-path Langevin dynamics for the $M$-dimensional system 
becomes 
\begin{eqnarray}
\dot{q}_{\alpha,k} &=&
G_{\alpha,k} -\Gamma_0 q_{\alpha,k} 
-\beta_T L_k \sum_{i=2}^{N-1} u_{ik} 
\left.
\frac{\partial V_{\rm eff}( \{ x_{\alpha} \})}{\partial x_{\alpha}} 
\right|_{x_{\alpha}=x_{\alpha,i}}
+\sqrt{2 L_k} \eta_{\alpha,k}(t),
\label{eq:FPL}
\end{eqnarray}
where
\begin{eqnarray}
\frac{\partial V_{\rm eff}( \{ x_{\alpha} \})}{\partial x_{\alpha}}
&=& 
\sum_{\beta=1}^{M}
\left[ \frac{1}{4 \omega_{\beta}^2} U_{x_{\beta}} U_{x_{\beta}x_{\alpha}}
-\frac{k_B T}{4 \omega_{\beta}^2} U_{x_{\beta} x_{\beta} x_{\alpha}}
\right],
\end{eqnarray}
and $L_k$ is chosen as in Eq.~(\ref{eq:Lk}) with $\omega=\omega_{\alpha}$.
Practically, the hessian and its derivative 
above are the bottleneck of this 
computation, hampering its application to large molecular systems.
Recently, however, the symmetric OM action,
which only uses the force to calculate the OM action, 
has been devised by Miller and Predescu \cite{MP07}.
Such a ``low cost'' action can be utilized for the future 
application of this method to real molecular systems.

\subsection{Replica exchange in path space}
\label{sec:OM-REM}

In the previous subsection, we have obtained a primitive way to 
generate a path ensemble by
solving the overdamped Langevin equation for Fourier components of a path.
However, path sampling efficiency becomes an issue 
in biomolecular simulations due to the extremely rugged-energy surface. 
In such a case, the sampled path often gets 
trapped around the initially assumed path, 
leading to unphysical results and incorrect predictions on the 
reaction mechanism. 
This situation is in parallel with the 
configuration sampling problem of complex molecules 
such as 
peptides or proteins
in a rugged-energy landscape. 

\begin{figure}
\hfill
\begin{center}
\includegraphics[scale=0.7]{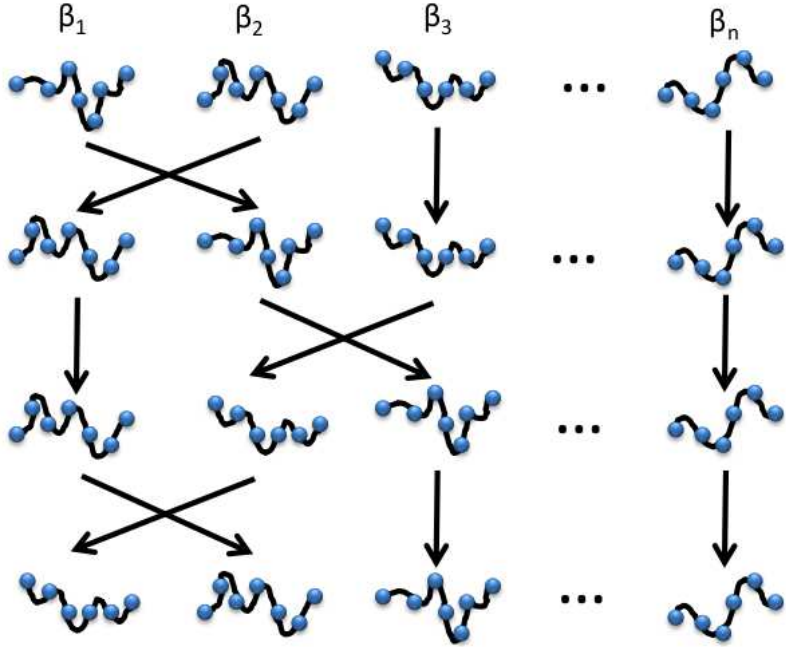}
\end{center}
\caption{\baselineskip5mm
Schematic drawing of the proposed replica-exchange method
for path sampling using the Onsager-Machlup action. 
Each column represents the OM dynamics of a single path.
Different path replicas with different inverse 
temperatures $\beta_i$ ($i=1,2, \cdots , n$) 
are exchanged according to the Metropolis criterion, 
Eqs.~(\ref{eq:REM1}) and (\ref{eq:REM2}).
}
\label{fig:OM-REM}
\end{figure}

Generalized ensemble methods such as multi-canonical
sampling \cite{NNK97,TMK03}
or replica exchange \cite{Hansmann97,SO99} (parallel tempering) are 
well-known to solve this problem. We here combine one of these methods,
the replica exchange method (REM), with the OM method 
because of its simplicity for implementation. 
It is easy to derive the 
following acceptance probability for exchange of two replicas
\begin{eqnarray}
P_{\rm acc} &=& \min (1,\exp \{\Delta \} ),
\label{eq:REM1}
\\
\Delta &=&
(\beta-\beta')({\cal H}-{\cal H}'),
\label{eq:REM2}
\end{eqnarray}
where $(\beta, {\cal H})$ and $(\beta', {\cal H}')$ are 
the inverse temperature 
and the OM Hamiltonian 
for each replica.
The difference with the conventional REM \cite{Hansmann97,SO99}
is that each replica is associated with a path (see Fig.~\ref{fig:OM-REM}),
and that Eq.~(\ref{eq:OMHam}) is employed for the effective 
Hamiltonian ${\cal H}$.
As usually done, we prepare several replicas for a path and,
during the Langevin dynamics in our case, 
exchange two ``neighboring'' paths
at the same time with the above probability \cite{SO99}.
If the number of replicas and the exchange rate 
between replicas are both sufficiently large to sample the 
whole ``path'' space,
an appropriate path ensemble with temperature $T$ 
should be obtained by collecting path data indexed 
by the same $T$.



\section{Results}
\label{sec:result}

\subsection{Model and methods}

\begin{figure}
\hfill
\begin{center}
\begin{minipage}{.42\linewidth}
\includegraphics[scale=1.4]{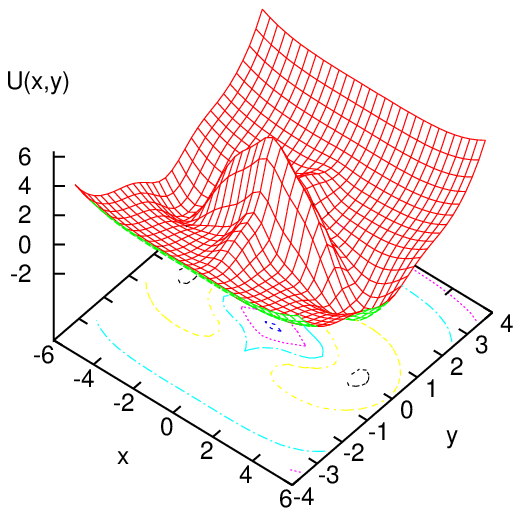}
\end{minipage}
\hspace{1cm}
\begin{minipage}{.42\linewidth}
\includegraphics[scale=1.2]{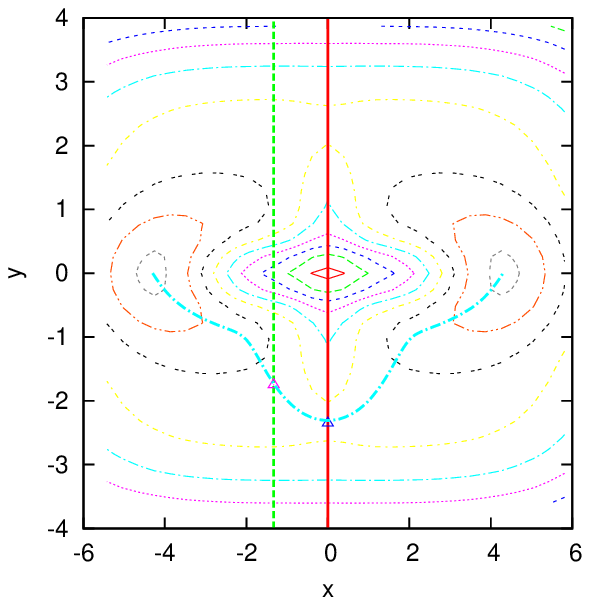}
\end{minipage}
\end{center}
\caption{\baselineskip5mm
Left: Bolhuis' potential defined by Eq.~(\ref{eq:bolhuis}) with 
$a=1$ and $b=0$ (symmetric case). There are two stable 
regions (basins) around $(x,y)=(\pm 4.3,0)$ and a 
large barrier separating two pathways connecting 
the two basins.
Right: Contour plot of the Bolhuis potential.
The red and green vertical lines represent the 
exit positions used later in the analysis. 
The blue curve connecting two minima represents
a minimum energy path and the intersections 
with the red and green lines are denoted as 
triangles.
}
\label{fig:Bolhuis}
\end{figure}

We test our method by using
Bolhuis' two-dimensional potential \cite{Bolhuis08} 
\begin{eqnarray}
U(x,y)
&=&
-3 e^{-0.25(x-4)^2-y^2}-3 e^{-0.25(x+4)^2-y^2}
\nonumber
\\
&&+\frac{32}{1800}(0.0625 x^4 +y^4) +5 e^{-0.0081x^4-4 y^2}
\nonumber
\\
&&
+2 e^{-1.5(x-b)^2-(y-1)^2} +2 a e^{-1.5(x+b)^2-(y+1)^2}.
\label{eq:bolhuis}
\end{eqnarray}
This potential has two minima at $(\pm 4.3, 0.0)$ 
which are interconnected by two reaction pathways
via two saddle points $(0.0,\pm 2.3)$.
These pathways are separated by a large energy barrier
centered at $(0.0,0.0)$. 
The parameters are taken as $a=1$ and $b=0$, resulting 
in a symmetric shape of the potential (see Fig.~\ref{fig:Bolhuis}).
This potential was chosen because it is a minimal model 
with multiple pathways, representing
the local character of the rugged energy landscape 
of biomolecules. 
Our aim is to sample
an ensemble of paths connecting two minima.

We need to determine the parameters for 
the overdamped Langevin equation, Eq.~(\ref{eq:Langevin}).
We first take $\gamma_{\alpha}=1$, i.e., the characteristic 
timescale of this Langevin dynamics is unity.
From the Einstein relation, Eq.~(\ref{eq:Einstein}), 
the diffusion constant
is equal to the thermal energy, $k_B T$.
Here we set $k_B T=1.25$, since this is an interesting case
where barrier crossing can easily occur in the $x$ direction,
but not in the $y$ direction.

Next we describe how to discretize a path in the Bolhuis 
potential.
The discretization of a path can be characterized by
the time interval $\Delta t$ and the number 
of the discretized path (the number of ``beads'') $N$,
i.e., the total time of the path is $t_{\rm tot}= N \Delta t$.
To estimate the reasonable value for $t_{\rm tot}$,
we consider free diffusion between the two minima with length $L$, 
where the diffusion timescale is $t_{\rm diff} = L^2/2D$.
In our case,
$L \simeq 2\times 4.3=8.6$ and $D=1.25$, and thus $t_{\rm diff} \simeq 30$.
However, if the potential surface is concave
in the transition state region, the motion can be quicker
than free diffusion, and
the actual time for barrier crossing
can be shorter than this estimate.
In this paper, we chose $t_{\rm tot}=12$.
The next question is how to choose $\Delta t$ or $N$ while $N \Delta t=12$.
After several trial and error, we found that $\Delta t=0.05$ and $N=240$ 
are appropriate for this Bolhuis potential.

We further need to determine the time step 
for the Fourier-path Langevin dynamics, Eq.~(\ref{eq:FPL}).
We used the simplest Euler scheme to solve this dynamics \cite{GSH88}, 
and found that 
$\lambda=10^5$ and 
$\Delta t_F =0.5$ are reasonable. Note that the 
Langevin dynamics for a path is carried out just for sampling and 
the timescale determined by these parameters have no physical meaning.
For comparison, we also numerically solve the 
original overdamped Langevin equation, Eq.~(\ref{eq:Langevin}), 
with the same Euler scheme.
We have confirmed that $\Delta t_L = 0.003$ is sufficient to 
recover the potential function through the formula 
$U(x,y) = -k_B T \log P(x,y)+C$, where $P(x,y)$ is the 
distribution function calculated by the Langevin trajectory
and $C$ is a constant.

Finally we describe how to choose the parameters for REM.
The most important prerequisite for REM is that the distribution 
of ${\cal H}$ 
should have sufficient overlap between neighboring 
replicas \cite{Hansmann97,SO99}, which is written as 
\begin{equation}
|\langle {\cal H} \rangle_{i+1} -\langle {\cal H} \rangle_i |
< 
\langle \Delta {\cal H} \rangle_{i+1} 
+\langle \Delta {\cal H} \rangle_i. 
\end{equation}
A rough estimate may be obtained from 
an independent particle 
approximation and equipartition principle.
For the spring contribution in Eq.~(\ref{eq:OMHam}),
the number of particles is approximately $N \times M$,
and the thermal energy $k_B T/2$ is given to each 
degree of freedom, the partial average value of the Hamiltonian ${\cal H}$
is therefore $NM k_B T/2$. Actually there is a contribution from 
the effective potential term $V_{\rm eff}$, which might be 
also approximated as an $N$-tiple connected harmonic spring, 
then the average of the total Hamiltonian may 
be $\langle {\cal H} \rangle = NM k_B T$.
Assuming that they are all uncorrelated Gaussian random numbers,
the fluctuations of the Hamiltonian 
becomes $\langle \Delta {\cal H} \rangle = \sqrt{2 NM}k_B T$.
Defining $f=T_{i+1}/T_i>1$, 
the criterion for the neighboring distributions to overlap is written as
\begin{equation}
NM(f-1)<\sqrt{2NM}(f+1),
\end{equation}
and we find 
\begin{equation}
1<f<1+\frac{2}{\sqrt{NM/2}-1}.
\end{equation}
In our case of the Bolhuis potential, 
$NM \simeq 480$, the right-hand side thus becomes 1.14.
In this paper, we prepare eight replicas by taking $f=1.14$, 
hence the corresponding temperatures 
are $k_B T= 3.60, 3.10, 2.66, 2.29, 1.97, 1.69, 1.46, 1.25$.
We have checked that the eight replicas are mixed up well 
and the equilibrium path distributions are sampled appropriately
at the respective temperatures.

\subsection{Numerical results}

\begin{figure}
\hfill
\begin{center}
\begin{minipage}{.42\linewidth}
\includegraphics[scale=1.2]{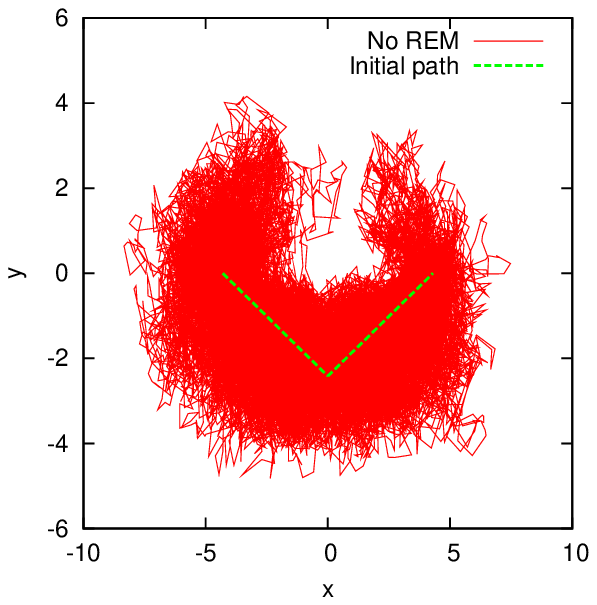}
\end{minipage}
\hspace{1cm}
\begin{minipage}{.42\linewidth}
\includegraphics[scale=1.2]{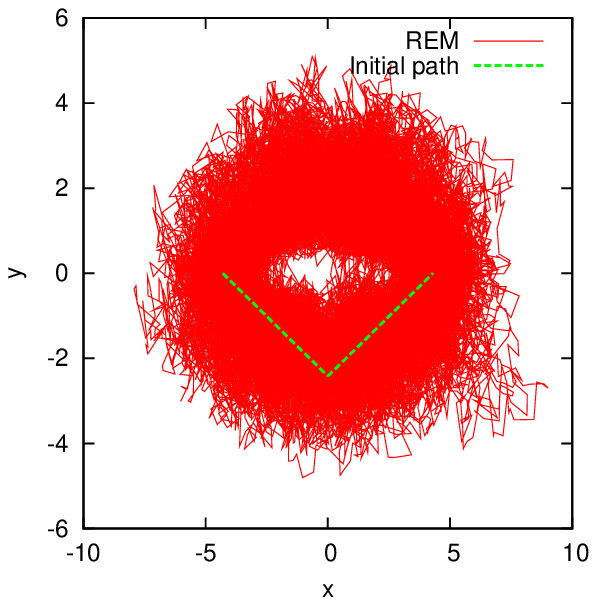}
\end{minipage}
\end{center}
\caption{\baselineskip5mm
Path ensemble generated by the Fourier-path dynamics
without REM (left) and with REM (right). 
The diffusion constant or temperature is $D=k_B T=1.25$. 
200 paths are superimposed for each panel with the interval of 
$30 \times \Delta t_F$. 
The initial guess is also drawn as a green line. 
}
\label{fig:OMtrj}
\end{figure}

In Fig.~\ref{fig:OMtrj},
we show the path ensembles generated by the 
Fourier-path dynamics with (right) or without (left) 
the use of REM.
An initially guessed path (green line in 
Fig.~\ref{fig:OMtrj}) was shifted from a linearly 
interpolated path to avoid the energy being too high.
Starting from the initial path,
the generated path gradually spreads out in path space.
However, in this ``timescale'' ($200 \times 30 \times \Delta t_F$), 
the path sampling without REM is not sufficient 
because the distribution turns 
out not to be symmetric
(see the left of Fig.~\ref{fig:OMtrj}).
This indicates that the simple Fourier-path 
dynamics does not converge fast enough when 
there are multiple pathways:
this is the situation of ``path trapping'' 
that we may encounter in path sampling simulations of biomolecules.
For such a case, REM should be employed to improve the sampling 
in path space.
As shown in the right of Fig.~\ref{fig:OMtrj},
the Fourier-path dynamics with REM 
can explore larger path space, resulting in a more symmetrical 
distribution as expected.

\begin{figure}
\hfill
\begin{center}
\begin{minipage}{.42\linewidth}
\includegraphics[scale=1.3]{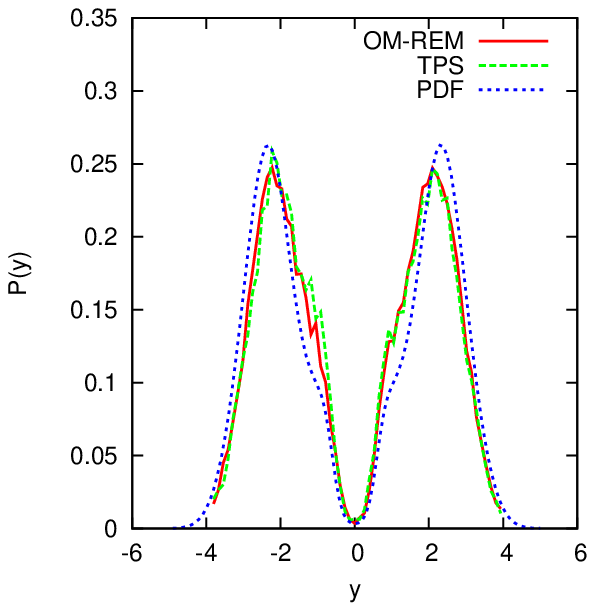}
\end{minipage}
\hspace{1cm}
\begin{minipage}{.42\linewidth}
\includegraphics[scale=1.3]{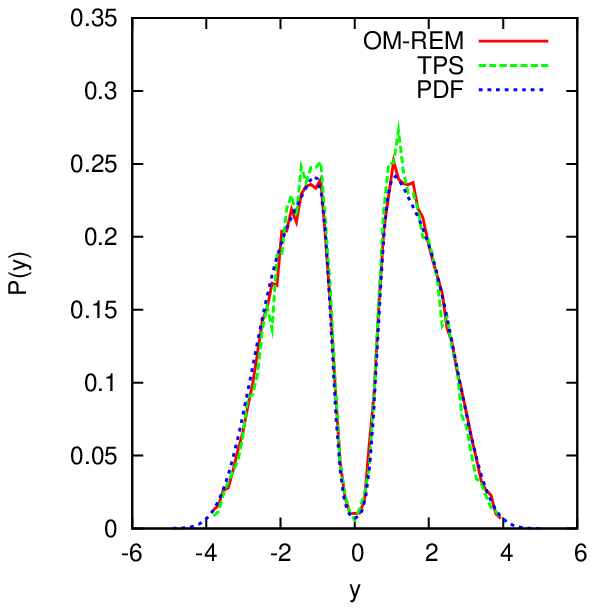}
\end{minipage}
\end{center}
\caption{\baselineskip5mm
Exit distributions on the transition state ($x_{\rm exit}=0$, left, 
corresponding to the red line on the right panel of Fig.~\ref{fig:Bolhuis})
and apart from the TS ($x_{\rm exit}=-1.33$, right, 
corresponding to the green line on the right panel of Fig.~\ref{fig:Bolhuis})
calculated by 
the OM method with REM (red) and TPS (green) with $D=k_BT=1.25$.
The equilibrium probability distribution function (PDF) 
is also shown (blue).
}
\label{fig:histgram}
\end{figure}

From the path ensemble thus generated,
we can obtain both nonequilibrium (such as kinetic rates) 
and equilibrium properties (such as free energy profiles).
Since we want to examine the nonequilibrium properties of 
the path ensemble,  
we employ an {\it exit distribution} \cite{CC01} 
as a measure of such nonequilibrium properties.
Note that the exit distribution is conceptually different 
from a free energy surface which is defined for 
equilibrium systems.
For our model system, the exit distribution is
defined as a histogram of accumulated $y$ 
when a trajectory first hits a cross section $x=x_{\rm exit}$.
In Fig.~\ref{fig:histgram}, we show 
the numerical result of the exit distributions 
for two exit positions: 
(a) one is on the transition state (TS):
$x_{\rm exit}=0$, 
corresponding to the red line 
in Fig.~\ref{fig:Bolhuis}, 
and 
(b) the other is apart from the TS: $x_{\rm exit}=-1.33$, 
corresponding to the green line 
in Fig.~\ref{fig:Bolhuis}. 
Because it 
is located on TS,
case (a) is expected to show stronger nonequilibrium 
behavior than case (b).
To check the correctness of our result, 
we also show the exit distributions 
calculated by transition path sampling (TPS) 
using the algorithm due to Crooks and Chandler \cite{CC01}.
For the TPS calculations,
the total time was $t_{\rm tot}=12$ as we used 
in the Fourier-path dynamics calculation, 
and a small cutoff distance (0.05) was introduced to 
define the reactant and product states 
around two minima.
We see that two exit distributions calculated by 
the two methods (at the different exit positions) agree
well within the statistical error.
This result indicates that our method using the 
OM action with REM can be alternative to TPS.
From computational points of view,
for this type of ``small'' system, TPS works best because 
it can easily sample the whole path space using the 
Crooks-Chandler algorithm.
The algorithm due to 
Eastman, Gronbech-Jensen, and Doniach \cite{EGD01}
is almost comparable to ours because both methods 
use the same information of the 
system such as the derivative of 
a hessian matrix.
Note, however, that our intention is to present
an efficient path sampling algorithm for large
systems. This approach of path sampling is considered 
to be more feasible for slow processes in large systems.

Next, the exit distributions are compared to 
the equilibrium probability function (PDF) at the exit points,
which is calculated as 
\begin{equation}
P_{\rm eq}(y) \propto \exp \{-\beta U(x_{\rm exit},y) \}
\label{eq:equidist}
\end{equation}
after normalization along the $y$ direction.
It is interesting to see that although the PDF 
is conceptually different from the exit distribution,
they look similar to each other at 
$x_{\rm exit}=-1.33$ (case (b), see the right of Fig.~\ref{fig:histgram}).
On the other hand,
the exit distribution is quite different from the 
PDF at 
$x_{\rm exit}=0.0$ (case (a), see the left of Fig.~\ref{fig:histgram}),
which should be attributed to nonequilibrium effects.
We have also computed the exit distributions 
using direct Brownian dynamics starting from 
one minimum (data not shown).
Interestingly, the result  
also gives a similar exit distribution for case (a) 
but not for case (b).
This indicates that the exit distribution does depend 
on the final state as well as the initial state, 
and for case (b) such an effect can be significant.
As such, our method or TPS should be carefully 
compared with direct Brownian dynamics.

\begin{figure}
\hfill
\begin{center}
\begin{minipage}{.42\linewidth}
\includegraphics[scale=1.3]{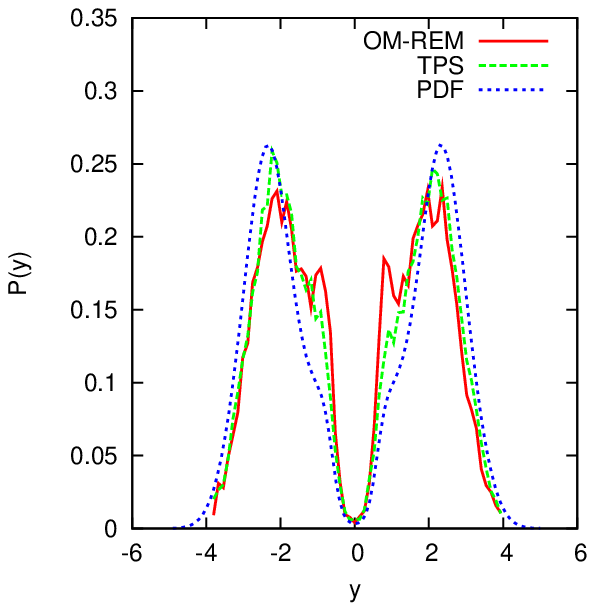}
\end{minipage}
\hspace{1cm}
\begin{minipage}{.42\linewidth}
\includegraphics[scale=1.3]{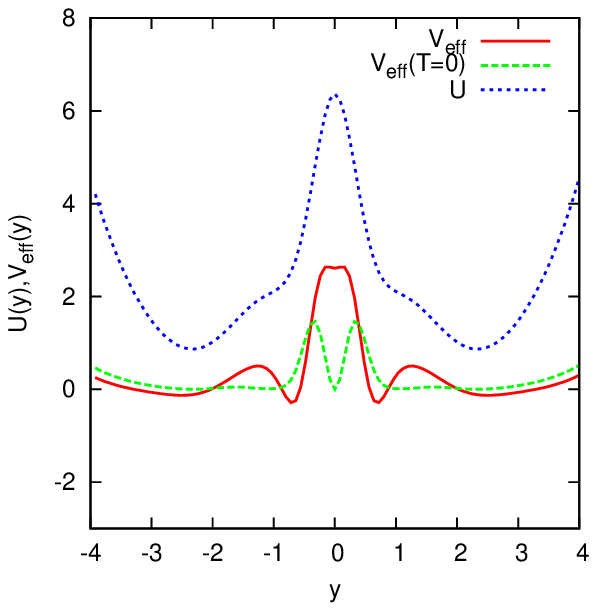}
\end{minipage}
\end{center}
\caption{\baselineskip5mm
Left: Exit distributions calculated by our method and  
TPS at the exit position 
$x_{\rm exit}=0$ with $k_B T=1.25$.
For the OM calculation,
$\Delta t=0.1$ and $N=120$ were chosen 
for the path disretization.
The probability distribution function 
at the exit position is also shown as a reference.
Right: Cross sections of the effective potential (red) 
and its zero-temperature component (green) at 
the exit position $x_{\rm exit}=0$ with $k_B T=1.25$. 
The cross section of the potential energy function 
is also shown in blue.
}
\label{fig:pitfall}
\end{figure}

The calculations done have been so far successful, 
and we now discuss practical aspects of the 
path sampling simulations.
In our approach, one of the important parameters that 
affect the computational effort is the length for 
path discretization $\Delta t$.
It is thus useful to mention how and why the calculation 
fails when we take larger $\Delta t$.
In the left of Fig.~\ref{fig:pitfall}, we show
the result of the exit distribution with 
$\Delta t =0.1$ and $N=120$ for case (a) while fixing the 
total time $t_{\rm tot}=12$ 
(the original setting was 
$\Delta t =0.05$ and $N=240$).
Compared with the left of Fig.~\ref{fig:histgram},
one can see that spurious peaks appear around $x= \pm 1.0$.

To understand these peak structures, we carefully examine the 
effective potential, Eq.~(\ref{eq:veff}). This potential 
can be decomposed into two parts: one is the zero-temperature 
component 
$V^{(1)}_{\rm eff}( \{ x_{\alpha} \})$
and the other the finite-temperature component
$V^{(2)}_{\rm eff}( \{ x_{\alpha} \})$,
that is, 
\begin{eqnarray}
V_{\rm eff}( \{ x_{\alpha} \})
&=&
V^{(1)}_{\rm eff}( \{ x_{\alpha} \})
+V^{(2)}_{\rm eff}( \{ x_{\alpha} \}),
\\
V^{(1)}_{\rm eff}( \{ x_{\alpha} \}),
&=&
\sum_{\alpha=1}^{M}
\frac{1}{8 \omega_{\alpha}^2} U_{x_{\alpha}}^2,
\\
V^{(2)}_{\rm eff}( \{ x_{\alpha} \})
&=&
-k_B T \sum_{\alpha=1}^{M}
\frac{1}{4 \omega_{\alpha}^2} U_{x_{\alpha}x_{\alpha}}.
\end{eqnarray}
It is easily seen that the finite-temperature component 
becomes zero when $T=0$.
More importantly, the finite-temperature component prefers a path 
with positive and large 
curvatures.
This is because an ensemble of trajectories becomes 
stable in the region where a curvature is large. (If the curvature 
is negative, the ensemble is unstable and spreading.)

In the right of Fig.~\ref{fig:pitfall}, we show the cross sections at the 
exit points for $V^{(1)}_{\rm eff}$ and 
$V_{\rm eff}$. From this figure, we can clearly see that 
the zero-temperature component does not have a sufficient ability 
to drag the path around $y \simeq \pm 1.0$.
We thus conclude that the peak structures in the exit distribution
stem from the finite-temperature 
component of the effective potential.
Hence we need to be careful when we 
deal with a path ensemble with large $\Delta t$ 
because some spurious features such as the 
peak structures might appear.


\section{Concluding remarks}
\label{sec:summary}


As a novel approach to sample diffusive paths based 
on the Onsager-Machlup action,
we have proposed Fourier-path Langevin dynamics.
To achieve powerful sampling in path space and to avoid the 
problem of ``path trapping'' around an initially guessed path, 
we have also suggested to combine this scheme 
with a powerful sampling technique, the replica 
exchange method.
Using the two dimensional model potential due 
to Bolhuis,
the validity of our method has been confirmed 
by the numerical comparison to the 
conventional transition path sampling.
We have also identified an interesting nonequilibrium path 
ensemble near the transition state of the model system, 
which is different from an equilibrium distribution.

We are now in a position to discuss the relation of our method with 
other path search or path sampling methods.
First, our main concern is the effects of temperature on a path, 
so the path search methods at a finite temperature 
such as MaxFlux methods \cite{HS97,CF03},
the temperature-dependent reaction coordinate \cite{ES00}, 
or finite-temperature string methods \cite{ERV05,VV09} 
have strong similarity with our method.
The difference is that 
a path generated by our method 
still holds nonequilibrium properties of the path
and there is no assumption such as 
 the existence of local equilibrium. 
It is interesting and important to investigate 
how the path ensembles calculated by different methods 
are actually different.

In this paper, for simplicity, we assumed overdamped Langevin dynamics, but 
this restriction can be easily relaxed. We can employ 
a modified action derived 
from the underdamped Langevin dynamics instead 
of the OM action \cite{MO53}. This strategy was successfully used for 
dynamic reweighting of a trajectory for a model system \cite{MWA08}.
When we apply this method to real molecular systems, 
we need to judge which Langevin dynamics (overdamped or underdamped) 
is more appropriate.

Of course, Eq.~(\ref{eq:odl}) is not the 
only dynamics that can generate a canonical 
ensemble of a path, 
e.g., one may use 
NVT molecular dynamics with 
N\'ose-Hoover thermostat techniques \cite{NH84,MKT92}.
In principle, a much simpler algorithm can be used 
such as a Monte Carlo (MC) move in path space.
That is, we devise a certain MC move $x \rightarrow x'$, and 
accept or reject the move using the Metropolis criterion \cite{Frenkelbook}.
However, we need a clever move when we apply the MC algorithm
to biomolecular systems, otherwise the move is rarely accepted.
The reason for this deficiency is the same as why a simple MC 
move does not work for sampling in biomolecular configuration space,
i.e., a MC move without consideration of a molecular configuration
causes a high energy state which will be rejected \cite{Kidera99}.
If we can devise a clever move in path space, it would be 
a powerful alternative to sample path space.
Combining the OM method with the other generalized 
ensemble methods such as the multi-canonical method \cite{NNK97,TMK03}
or the Tsallis ensemble method \cite{AS97,KS09} is also possible and promising.



We are grateful to Artur Adib 
for sending us a TPS code, 
and to Yuji Sugita, Tohru Terada, Ikuo Fukuda, 
Kei Moritsugu, Mikito Toda, 
and Hiroshi Teramoto for useful discussions.
This research was supported by Research and Development of 
the Next-Generation Integrated Simulation of Living Matter,
a part of the Development and Use of the Next-Generation 
Supercomputer Project of the Ministry of Education,
Culture, Sports, Science and Technology (MEXT).


\end{document}